\documentclass[11pt]{article}
\usepackage{graphicx}

\newcommand{\B}{B^1}
\newcommand{\f}{f^1}
\newcommand{\uR}{u^1_R}
\newcommand{\dR}{d^1_R}
\newcommand{\qR}{q^1_R}
\newcommand{\lR}{\ell^1_R}

\def\singleandabitspaced{\baselineskip=\normalbaselineskip\multiply
    \baselineskip by 110\divide\baselineskip by 100}
\def\singlespaced{\baselineskip=\normalbaselineskip}

\newcommand{\ra}{\rightarrow}

\newcommand{\centeron}[2]{{\setbox0=\hbox{#1}\setbox1=\hbox{#2}\ifdim
                             \wd1>\wd0\kern.5\wd1\kern-.5\wd0\fi \copy0
                             \kern-.5\wd0\kern-.5\wd1\copy1\ifdim\wd0>\wd1
                             \kern.5\wd0\kern-.5\wd1\fi}}
\newcommand{\ltap}{\>\centeron{\raise.35ex\hbox{$<$}}
                     {\lower.65ex\hbox{$\sim$}}\>}
\newcommand{\gtap}{\>\centeron{\raise.35ex\hbox{$>$}}
                     {\lower.65ex\hbox{$\sim$}}\>}
\newcommand{\gsim}{\mathrel{\gtap}}
\newcommand{\lsim}{\mathrel{\ltap}}

\begin{document}

\singlespaced

\begin{titlepage}

\begin{flushright}
hep-ph/0208261 \\
MADPH-02-1304 \\
\end{flushright}

\begin{center}
\vspace*{0.8in}
\mbox{\Large \textbf{Probing Kaluza-Klein Dark Matter 
with Neutrino Telescopes}} \\
\vspace*{1.6cm}
{\large Dan Hooper and Graham D. Kribs} \\
\vspace*{0.5cm}
{\it Department of Physics, University of Wisconsin, 
Madison, WI, USA 53706} \\
\vspace*{0.6cm}
{\tt hooper@pheno.physics.wisc.edu, kribs@physics.wisc.edu} \\
\vspace*{1.5cm}
\end{center}

\begin{abstract} 
\singleandabitspaced

In models in which all of the Standard Model fields live in extra
``universal'' dimensions, the lightest Kaluza-Klein (KK) particle 
can be stable.  Calculations of the one-loop radiative corrections 
to the masses of the KK modes suggest that the identity of the 
lightest KK particle (LKP) is mostly the first KK excitation of the 
hypercharge gauge boson.  This LKP is a viable dark matter candidate 
with an ideal present-day relic abundance if its mass is 
moderately large, between $600$ to $1200$ GeV\@.  
Such weakly interacting dark matter particles are expected to become 
gravitationally trapped in large bodies, such as the Sun, and 
annihilate into neutrinos or other particles that decay into neutrinos.
We calculate the annihilation rate, neutrino flux and the resulting event
rate in present and future neutrino telescopes.  The relatively large mass 
implies that the neutrino energy spectrum is expected to be well above 
the energy threshold of AMANDA and IceCube.  We find that the
event rate in IceCube is between a few to tens of events per year.
\end{abstract}

\end{titlepage}

\newpage
\setcounter{page}{2}
\singleandabitspaced

\section{Introduction}

The premiere astrophysical conundrum is the nature and identity
of dark matter.  The accumulated body of evidence in favor of the
existence of dark matter is by now overwhelming:
Studies of the cosmic microwave background \cite{cmb}, high redshift
supernovae \cite{sn}, galactic clusters and galactic rotation curves
\cite{rotation} indicate that the matter density of the universe
is $\Omega_M \simeq 0.3-0.4$. Constraints from big-bang nucleosynthesis,
however, limit the baryonic matter density to a small fraction of this
number \cite{bbn}.  Furthermore, the observed density of luminous matter
is also very small, $\Omega_L < 0.01$ \cite{luminous}.  Therefore, the
vast majority of the mass in the universe is dark.  Additionally, cosmic
microwave background studies and large scale structure formation requires
that the majority of the dark matter be cold (non-relativistic)
\cite{cmb,structure}.

Dark matter could exist in several forms.  
Perhaps the most interesting
possibility is a neutral, stable, weakly interacting particle
arising from physics beyond the Standard Model.  Candidates for such
an animal abound, including the lightest supersymmetric particle, 
the axion, etc.  The possibility of a stable Kaluza-Klein (KK) excitation 
as particle dark matter was raised many years ago \cite{KolbSlansky}
and more recently \cite{DDG2}.  Models in which all of the Standard
Model fields propagate in ``universal'' extra dimensions \cite{ACD}
(for earlier work, see \cite{antoniadis})
provide the most natural home for KK dark matter \cite{CMS,ST,CFM}.
This is because bulk interactions do not violate higher 
dimensional momentum conservation (KK number), and in these models
all of the couplings among the Standard Model particles arise from
bulk interactions.  To generate chiral 
fermions at the zero mode level, the extra compact dimension(s) must be  
modded out by an orbifold.  For five dimensions this is $S^1/Z_2$, 
while in six dimensions $T^2/Z_2$ is suitable and has other interesting 
properties \cite{ACD} including motivation for three generations
from anomaly cancellation \cite{DP} and the prevention of fast
proton decay \cite{ADPY}.  An orbifold does, however, lead to some
of the less appealing aspects of the model.  Brane-localized
terms can be added to both orbifold fixed points that violate KK number.
If these brane localized terms are symmetric under the exchange
of the two orbifold fixed points, then a remnant of KK number 
conservation remains, called KK parity.  All odd-level KK modes
are charged under this discrete symmetry thereby ensuring that the lightest 
level-one KK particle (LKP) does not decay.  This is entirely analogous
to exactly conserved $R$-parity in supersymmetric models which ensures
the lightest supersymmetric particle is stable.  The stability of the LKP 
suggests it could well be an interesting dark matter candidate.

The identity of the lightest KK particle crucially depends on
the mass spectrum of the first KK level.  At tree-level, the 
mass of each excitation is simply
\begin{equation}
\left(m^1_i\right)^2 = \frac{1}{R^2} + \left(m^0_i\right)^2
\end{equation}
where $R$ is the compactification radius that could be as large
as $1/(300 \; \mathrm{GeV})$ without conflict with experiment \cite{ACD}.
However, brane-localized terms can be added on the orbifold fixed
points that significantly modify the masses and higher 
dimensional wavefunctions.  The tree-level (matching) contributions
at the cutoff scale of the higher dimensional theory are not
calculable, but can be estimated using naive dimensional analysis.
The result is that the size of these terms are suppressed by a
volume factor, of order $\Lambda R$ where $\Lambda$ is the cutoff.
More importantly, these brane-localized terms are renormalized upon 
evolving from the matching scale to the mass scale of the 
light KK modes \cite{GGH}.  For universal extra dimensions, 
Cheng, Matchev, and Schmaltz showed that the tree-level mass 
formula receives significant one-loop radiative corrections
from log-enhanced brane-localized terms on the orbifold fixed 
points \cite{CMS}.
These radiative corrections are, in many cases, larger than the shifts 
in the tree-level masses resulting from the masses of the zero modes.
Indeed, here we will generally assume that these contributions
dominate over the tree-level volume-suppressed matching contributions.
Then, with the further assumption that the KK excitation of the Higgs 
does not receive a (significant) brane-localized negative contribution 
to its mass, the identity of the lightest KK state is identified as the 
first KK excitation of the photon.  Like the ordinary photon, the 
KK photon is an admixture between the first KK hypercharge gauge boson 
and the first KK neutral SU(2) gauge boson.  However, this identification 
is somewhat misleading, as \cite{CMS} point out, since the mixing angle 
for the level-one KK gauge bosons is generally much smaller than the 
Weinberg angle.  A leading order approximation to the mass of the 
lightest $\B$ state is 
\begin{equation}
m_{\B}^2 \simeq \frac{1}{R^2} \left[ 1 + \frac{g'^2}{16 \pi^2} \left( 
- \frac{39 \zeta(3)}{2 \pi^2} - \frac{1}{3} \ln \Lambda R + (2 \pi R v)^2 
\right) \right]
\end{equation}
neglecting higher order ${\cal O}(\sin \theta^1 v^2)$ corrections.
This approximation is equivalent to identifying 
LKP $\equiv \gamma^1 \simeq \B$, which we do for the 
remainder of the paper.

The relic density of the $\B$ has been calculated in 
a recent paper by Servant and Tait \cite{ST}.  Assuming the
LKPs were once in thermal equilibrium, they found that the
relic density is in the favorable region for providing the cold dark 
matter of the universe, $\Omega_{\B} h^2 = 0.16 \pm 0.04$, when the mass 
is moderately heavy, between $600$ to $1200$ GeV\@.
The range of mass depends on the relative importance of 
coannihilation with KK modes near in mass to the LKP\@.
We shall see later that coannihilation is active throughout
the parameter space when the KK mass spectrum is obtained using 
the radiative corrections arising from renormalized brane-localized terms.
Direct detection of the LKP as dark matter has been considered 
by Cheng, Feng, and Matchev \cite{CFM}.  They emphasized that
unlike the case of a neutralino LSP, the bosonic nature of the LKP 
means there is no chirality suppression of the annihilation signal
into fermions.  The annihilation rate of the LKP is therefore roughly 
proportional to the (hypercharge)$^4$ of the final state, leading 
to a large rate into leptons.  The large mass of the LKP suggests
that dark matter detection experiments sensitive to very heavy mass
relics should be among the most promising methods of 
detection.  In this paper, we explore the signal expected at present 
and future high energy neutrino telescopes that can probe precisely
this type of heavy dark matter.\footnote{Note that \cite{ST,CFM} 
also remarked on the potential importance of indirect detection 
at neutrino telescopes.}

\section{Relic density and KK mass spectrum}

The relic density of $\B$'s depends on the $\B$ mass, the annihilation
cross section, and the coannihilation rate.  The annihilation and
coannihilation cross sections are determined by SM couplings and the 
mass spectrum of the first KK level.  In \cite{ST}, the relic density
of $\B$'s was calculated in the following approximation: the masses
of all of the first level KK modes are equal, and electroweak breaking
effects (i.e., fermion and gauge boson masses) are neglected.  
In the low-velocity limit their result is
\begin{equation}
\langle \sigma v \rangle = \frac{95 g_1^4}{324 \pi m_{\B}^2} \; ,
\label{relic-density-eq}
\end{equation}
into fermion final states.  There are also annihilations into Higgs boson
pairs, but this is only a few percent additive correction to the above 
and so can be neglected.  A more general calculation for arbitrary
masses of intermediate state KK fermions is straightforward, but
as we will see, not necessary to obtain the quantitative results
we present below.

The full KK mass spectrum with one-loop corrections is presented in
\cite{CMS}.  The general expression for the correction to the level one
KK fermion masses is given by
\begin{equation}
m_{\f} = \frac{1}{R} \left[ 1 + \frac{9}{32 \pi^2} \ln \Lambda R \, 
\sum_a C_a(f) g_a^2 \right] \; .
\end{equation}
(The Yukawa corrections to this formula can be safely ignored 
since the KK top plays a negligible role in the analysis to follow.)
The sum is over all of the SM gauge groups, and $C_a(f)$ is the
quadratic Casimir for the fermion transforming under the group $a$.
The scale $\Lambda$ is the strong coupling cutoff scale of the 
extra dimensional theory.  
An important observation from this general formula is that the
fractional shift in the mass to the KK quarks is generally an 
order of magnitude larger than for the KK leptons.  This is simply
due to the much larger QCD corrections over the electroweak 
corrections.  In particular, consider the following fractional shifts
\begin{eqnarray}
r_{\lR} &\equiv& \frac{m_{\lR} - m_{\B}}{m_{\B}} = 
\frac{g'^2}{32 \pi^2} \left[ \frac{28}{3} \ln \Lambda R 
+ \frac{39 \zeta(3)}{2 \pi^2} - (2 \pi R v)^2 \right] \label{lR-eq} \\
r_{\uR} &\equiv& \frac{m_{\uR} - m_{\B}}{m_{\B}} = 
\frac{6 g_3^2}{16 \pi^2} \ln \Lambda R +
\frac{g'^2}{32 \pi^2} \left[ \frac{13}{3} \ln \Lambda R 
+ \frac{39 \zeta(3)}{2 \pi^2} - (2 \pi R v)^2 \right] \; . \label{uR-eq} \\
r_{\dR} &\equiv& \frac{m_{\dR} - m_{\B}}{m_{\B}} = 
\frac{6 g_3^2}{16 \pi^2} \ln \Lambda R +
\frac{g'^2}{32 \pi^2} \left[ \frac{4}{3} \ln \Lambda R 
+ \frac{39 \zeta(3)}{2 \pi^2} - (2 \pi R v)^2 \right] \; . \label{dR-eq}
\end{eqnarray}
The non-logarithmically enhanced terms contribute at most
a percent (for $1/R = 300$ GeV), and are negligible for the
radii of interest here ($1/R \gsim 600$ GeV).  The ratio of
the fractional shifts is
\begin{equation}
\frac{r_{\qR}}{r_{\lR}} \simeq \frac{18 g_3^2}{14 g'^2} \simeq 12 \qquad
\mathrm{where} \qquad r_{\qR} \simeq r_{\uR} \simeq r_{\dR} \; .
\label{ratio-eq}
\end{equation}
Notice that the ratio is independent of the compactification radius
and also independent of the cutoff scale.  Hence, the relative KK 
particle mass hierarchy is fixed by the just the SM couplings.  

What is the size of the radiative correction for a given KK mode?  
This is dependent on the cutoff scale in theory.  Strong coupling in 
extra dimensional theories appears when $g^2 N$ is order one, 
where $g$ is the coupling and $N$ is the number of (KK) particles 
exchanged.  For five universal dimensions this has been estimated 
to be $\Lambda \simeq 20/R$ \cite{ACD}.  This results in the fractional 
shifts $r_{\lR} \simeq 0.011$ and $r_{\qR} \simeq 0.14$.  For six 
universal dimensions the fractional shifts are about half as big.  

Several observations based on this discussion of the KK mass
spectrum are in order:  First, the right-handed
KK lepton is within a percent or so of the $\B$ mass, implying
that coannihilation of the $\B$ with $\lR$ is always 
important.  This means that for the $\B$ to have the 
appropriate abundance to be the dark matter of the Universe, it 
must have a mass between about $600$ to $800$ GeV using the 
results of \cite{ST}.  Second, we see that the estimate of the 
relic density (\ref{relic-density-eq}) in which all of the level one 
KK modes were taken to have the same mass is a reasonably good 
approximation since none of the KK modes is more than about 
$20\%$ heavier than the $\B$.

\section{Capture and annihilation in the Sun}

The calculation of the flux of neutrinos from WIMP annihilations in the
Sun (and Earth) has been explored in some detail, especially for the case
of neutralino dark matter \cite{susydm}.  The basic idea is to 
begin with the relatively well-known local dark matter density from
the galactic rotation data, compute the interaction cross section 
of the WIMPs with nuclei in the Sun, compare the capture rate 
with the annihilation rate to determine if these processes
are in equilibrium, and then compute the flux of neutrinos that
result from this rate of WIMP capture and annihilation.

There are two separate channels by which WIMPs can scatter off
nuclei in the Sun:  spin-dependent interactions and spin-independent
interactions.  For accretion from spin-dependent scattering, the capture 
rate is \cite{capture} 
\begin{equation} 
C_{\mathrm{SD}}^{\odot} \simeq 3.35 \times 10^{18} \, \mathrm{s}^{-1} 
\left( \frac{\rho_{\mathrm{local}}}{0.3\, \mathrm{GeV}/\mathrm{cm}^3} \right) 
\left( \frac{270\, \mathrm{km/s}}{\bar{v}_{\mathrm{local}}} \right)^3  
\left( \frac{\sigma_{\mathrm{H, SD}}} {10^{-6}\, \mathrm{pb}} \right)
\left( \frac{1000 \, \mathrm{GeV}}{m_{\B}} \right)^2 
\label{c-eq}
\end{equation} 
where $\rho_{\mathrm{local}}$ is the local dark matter density, 
$\sigma_{\mathrm{H,SD}}$ is the spin-dependent, WIMP-on-proton (hydrogen)
elastic scattering cross section, $\bar{v}_{\mathrm{local}}$ 
is the local rms velocity of halo dark matter particles and 
$m_{\B}$ is our dark matter candidate.  The analogous formula for the 
capture rate from spin-independent (scalar) scattering is \cite{capture}
\begin{equation}
C_{\mathrm{SI}}^{\odot} \simeq 1.24 \times 10^{18} \, \mathrm{s}^{-1} 
\left( \frac{\rho_{\mathrm{local}}}{0.3\, \mathrm{GeV}/\mathrm{cm}^3} \right) 
\left( \frac{270\, \mathrm{km/s}}{\bar{v}_{\mathrm{local}}} \right)^3 
\left( \frac{2.6 \, \sigma_{\mathrm{H, SI}}
+ 0.175 \, \sigma_{\mathrm{He, SI}}}{10^{-6} \, \mathrm{pb}} \right) 
\left( \frac{1000 \, \mathrm{GeV}}{m_{\B}} \right)^2 \; .
\end{equation}
Here, $\sigma_{\mathrm{H, SI}}$ is the spin-independent, WIMP-on-proton
elastic scattering cross section and $\sigma_{\mathrm{He, SI}}$ is the 
spin-independent, WIMP-on-helium elastic scattering cross section. 
Typically, $\sigma_{\mathrm{He, SI}} \simeq 16.0 \, \sigma_{\mathrm{H, SI}}$.
The factors of $2.6$ and $0.175$ include information on the solar 
abundances of elements, dynamical factors and form factor suppression.

Although these two rates appear to be comparable in magnitude, 
the spin-dependent cross section is typically three to 
four orders of magnitude larger than the spin-independent 
cross section \cite{CFM}.  Therefore, solar accretion by spin-dependent 
scattering dominates, and hereafter we can safely ignore 
spin-independent scattering.
The microscopic elastic scattering spin-dependent cross section of 
a $\B$ off of a proton was calculated in \cite{CFM}.  Their result 
can be well-approximated by
\begin{equation}
\sigma_{\mathrm{H,SD}} = \frac{g'^4 m_p^2}{648 \pi m_{\B}^4 r_{\qR}^2} 
\left( 4 \Delta_u^p + \Delta_d^p + \Delta_s^p \right)^2
\end{equation}
since the scattering cross section is dominated by exchanges of 
right-handed KK quarks due to their larger hypercharge over 
left-handed KK quarks.  Here, $m_p$ is the mass of the proton and 
the $\Delta^p_q$'s parameterize the fraction of spin carried by 
a constituent quark $q$ \cite{spinfraction},
\begin{equation}
\Delta_u^p = 0.78 \pm 0.02 \quad , \quad 
\Delta_d^p = -0.48 \pm 0.02 \quad , \quad 
\Delta_s^p = -0.15 \pm 0.07 \; .
\end{equation}
Inserting the spin-fractions, we obtain
\begin{equation}
\sigma_{\mathrm{H,SD}} = 0.9 \times 10^{-6} \, \mathrm{pb} 
\left( \frac{1000 \, \mathrm{GeV}}{m_{\B}} \right)^4 
\left( \frac{0.14}{r_{\qR}} \right)^2 \; .
\end{equation}

If the capture rates and annihilation cross sections are sufficiently
high, the Sun will reach equilibrium between these processes.  
For $N$ number of $\B$'s in the Sun, the rate of change of this
number is given by
\begin{equation}
\dot{N} = C^{\odot} - A^{\odot} N^2 \; ,
\end{equation}
where $C^{\odot}$ is the capture rate and $A^{\odot}$ is the 
annihilation cross section times the relative WIMP velocity per volume.  
$C^{\odot}$ was given in (\ref{c-eq}), while $A^{\odot}$ is 
\begin{equation}
A^{\odot} = \frac{\langle \sigma v \rangle}{V_{\mathrm{eff}}} 
\end{equation}
where $V_{\mathrm{eff}}$ is the effective volume of the core
of the Sun determined roughly by matching the core temperature with 
the gravitational potential energy of a single WIMP at the core
radius.  This was found in \cite{equ} to be
\begin{equation}
V_{\rm eff} = 1.8 \times 10^{26} \, \mathrm{cm}^3 
\left( \frac{1000 \, \mathrm{GeV}}{m_{\B}} \right)^{3/2} \; .
\end{equation}
The present $\B$ annihilation rate is 
\begin{equation} 
\Gamma = \frac{1}{2} A^{\odot} N^2 = \frac{1}{2} \, C^{\odot} \, 
\tanh^2 \left( \sqrt{C^{\odot} A^{\odot}} \, t_{\odot} \right) \; 
\end{equation}
where $t_{\odot} \simeq 4.5$ billion years is the age of the solar system.
The annihilation rate is maximized when it reaches equilibrium with
the capture rate.  This occurs when 
\begin{equation}
\sqrt{C^{\odot} A^{\odot}} t_{\odot} \gg 1 \; .
\end{equation}
Combining our expression for the capture and annihilation rate
[using $\langle \sigma v \rangle$ from (\ref{relic-density-eq})],
we find
\begin{equation}
\sqrt{C^{\odot} A^{\odot}} \, t_{\odot} = 
2.4 \left( \frac{1000 \, \mathrm{GeV}}{m_{\B}} \right)^{13/4} 
\frac{0.14}{r_{\qR}} \; .
\end{equation}
Hence, throughout the mass range that leads to the ideal relic
abundance of $\B$'s considered by \cite{ST}, we find that the Sun 
either reaches or nearly reaches equilibrium between $\B$ capture
and annihilation.

The Earth, being less massive and accreting dark matter only by scalar
interactions, captures particles much more slowly than the Sun.  For the
optimistic case of $m_{\B} = 500$ GeV with $\sigma_{\mathrm{H}}\sim
10^{-6}$ pb (scalar only), the ratio of the age of the solar
system to the equilibrium time is on the order of $10^{-5}$, which
corresponds to a $10^{-10}$ suppression in the annihilation rate.  
We find no scenario in which KK dark matter annihilation in the 
Earth provides an observable signal.

\section{Event rates and prospects for detection}

Now that we have determined the annihilation rate in the Sun,
we need to determine the outgoing flux of detectable particles.
This corresponds to determining the annihilation fraction
directly into muon neutrinos, as well as indirectly through
decays.  We are interested exclusively in muon neutrinos since
at the energies relevant to $\B$ annihilation, neutrino telescopes
only observe muon tracks generated in charged-current interactions.
In the approximation that all heavier level one KK modes
have the same mass, the relative annihilation fraction can
be determined from simply the hypercharge of the final state
fermions.  This is shown in Table~\ref{BR-table} for the 
column $r_{\f} = 0$.  
\begin{table}[t]
\begin{center}
\renewcommand{\arraystretch}{1.2}\small\normalsize
\begin{tabular}{rcl|ll} \hline\hline
\multicolumn{3}{c}{process} & \multicolumn{2}{|c}{annihilation fraction} \\
& & & $r_{\f} = 0$ & $r_{\qR} = 0.14$ \\ \hline
$\B\B$ & $\ra$ & $\nu_e \overline{\nu}_e$, $\nu_\mu \overline{\nu}_\mu$,
                 $\nu_\tau \overline{\nu}_\tau$ & $0.012$ & $0.014$ \\
       & $\ra$ & $e^+e^-$, $\mu^+\mu^-$, 
                 $\tau^+\tau^-$ & $0.20$ & $0.23$ \\
       & $\ra$ & $u\overline{u}$, $c\overline{c}$, 
                 $t\overline{t}$ & $0.11$ & $0.077$ \\
       & $\ra$ & $d\overline{d}$, $s\overline{s}$, 
                 $b\overline{b}$ & $0.007$ & $0.005$ \\ 
       & $\ra$ & $\phi \phi^*$
                 & $0.023$ & $0.027$ \\ \hline\hline
\end{tabular}
\end{center}
\caption{The relative annihilation fraction into various 
final states.  The numbers shown are not summed over generations,
and the Higgs mass was assumed to be lighter than $m_{\B}/2$.}
\label{BR-table}
\end{table}
However, the annihilation into KK quarks is slightly further
suppressed since the KK quarks are slightly heavier than the KK leptons.
Using $r_{\qR} = 0.14$, our estimates of the modified branching fractions 
are shown in Table~\ref{BR-table}.
Clearly the relative annihilation fractions
are not particularly sensitive to the details of the spectrum
so long as KK leptons are the same mass or lighter than KK quarks 
as the one-loop radiative corrections suggest.

Neutrinos are generated in annihilations directly, but can also be
produced in the decays of tau leptons, quarks and Higgs bosons generated
in annihilations.  Only very short lived particles contribute to secondary
neutrino flux, as longer lived particles lose the majority of their energy
from scattering in the Sun before decaying.  Bottom and charm quarks lose
energy from hadronization before decaying. Although top quarks do not
hadronize, they generate neutrinos only through gauge bosons and bottom
quarks generated in their decay.  Neutrinos can also lose energy traveling
through the Sun \cite{edsjo,JK,crotty}.  Muon neutrinos that escape the
Sun then travel to the Earth where a small fraction of them are converted 
to muons through charged current interactions.  Neutrino telescopes 
observe high energy muon neutrinos by identifying a muon track in the 
detector medium generated in a charged-current interaction (for a review, 
see \cite{review}).

To calculate the event rate in neutrino telescopes, we first must 
calculate the spectrum of neutrinos from annihilation directly to 
neutrinos as well as from other annihilation modes that lead indirectly
to neutrinos from decays.  Calculating the flux of indirectly produced 
neutrinos requires modeling the physics of the final state, including 
hadronization and decay, and must be done numerically.  Fortunately, 
the indirect sources have been previously calculated in supersymmetric
models for neutralino annihilation in the Sun.  The flux of muon 
neutrinos from the indirect final states of leptons, quarks, 
gauge bosons, and Higgs particles has been simulated and parameterized 
by Edsjo \cite{edsjo}.  We use this parameterization, replacing
the neutralino with the LKP, and using the relative annihilation
rates into the various final states as shown in Table~\ref{BR-table}.
In this parameterization the flux of muons expected at an Earth-based 
detector is
\begin{equation}
\label{edsjo}
\frac{d \Phi_\mu}{d z}(m_{\B},z) = \frac{p_1 m_{\B}^2 
\left[ 1 - \exp\left( -p_5 \frac{m_{\B}}{z^{p_8}} \right) \right]
\exp\left( -p_7 m_{\B} z \right)}{1 + 
\exp\left( \frac{z - p_2 \exp(- p_3 m_{\B}) - p_6}{p_4} \right)}
\label{para-eq}
\end{equation}
where $z \equiv E_\mu/m_{\B}$ and $p_1 \ldots p_8$ are the parameters
fitted to the simulation results that can be found in \cite{edsjo}.
The accuracy of this parameterization was estimated to be within about 
15\% so long as the threshold is not too close to the LKP mass, 
$E^{\rm th}_\mu \lsim 0.2 \, m_{\B}$.  We have used a muon energy 
threshold of $50$ GeV, although our results are not very sensitive 
to this choice.  This energy threshold is well below the masses
of the LKPs considered here, and thus well within the range where this
parameterization is expected to be valid.

The calculation of the direct neutrino signal was done separately.
Annihilation of LKPs into neutrinos results in a mono-energetic 
spectrum of neutrinos.  Of course not all of the neutrinos escape
the Sun, due to interactions in the solar medium.  The probability of 
a neutrino escaping the Sun without interacting is given by \cite{crotty}
\begin{equation} 
P = e^{-E_{\nu}/E_k}
\end{equation}
where $E_k \simeq$ ($130$, $160$, $200$, $230$) GeV for 
($\nu_\mu$, $\nu_\tau$, $\overline{\nu}_\mu$, $\overline{\nu}_\tau$). 
We make the conservative approximation that all neutrinos that interact 
in the Sun are absorbed rather than regenerated.  Note that this 
exponential suppression of very high energy neutrinos emitted from
the Sun is implicitly included in the parameterization, Eq.~(\ref{para-eq}).

Neutrinos exiting the Sun travel to the Earth and then occasionally
interact through charged current interactions.  The mono-energetic spectrum 
of muon neutrinos becomes a distribution of muon energies \cite{crosssection} 
due to the conversion interactions.  The resulting muon travels a
distance $R_{\mu}$ before its energy falls below the threshold energy 
$E_{\mathrm{th}}$.  This distance, called the muon range, is given 
by \cite{range}
\begin{equation} 
R_{\mu} \simeq \frac{1}{\rho \beta} \mathrm{ln}\bigg[\frac{\alpha + \beta E_{\mu}}{\alpha+\beta E_{\mathrm{th}}} \bigg]
\end{equation}
where $\rho$ is the density of the detector medium, 
$\alpha \simeq 2.0$ MeV cm$^2$/g and 
$\beta \simeq 4.2 \times 10^{-6}$ cm$^2$/g. 
The effective volume in which a muon producing interaction can occur 
and be observed is simply the muon range times the effective area of 
the detector.  

Another important effect for all sources of neutrinos is flavor 
oscillations.  In the Sun, for the energies of interest to us here, 
an originally outgoing muon or tau neutrino oscillates sufficiently 
so as to randomize its final flavor \cite{crotty}.  (Oscillation into 
electron neutrinos occurs only for a significantly lower energy source of
neutrinos.)  This gives a significant increase 
to the number of muon neutrinos that result from the annihilation
process $\B\B \ra \tau^+\tau^-$, since every $\tau$ decay results
in a tau neutrino that has about an even chance of exiting the Sun 
as a muon neutrino.

\begin{figure}[t]
\centering\leavevmode
\mbox{
\includegraphics[width=3.2in]{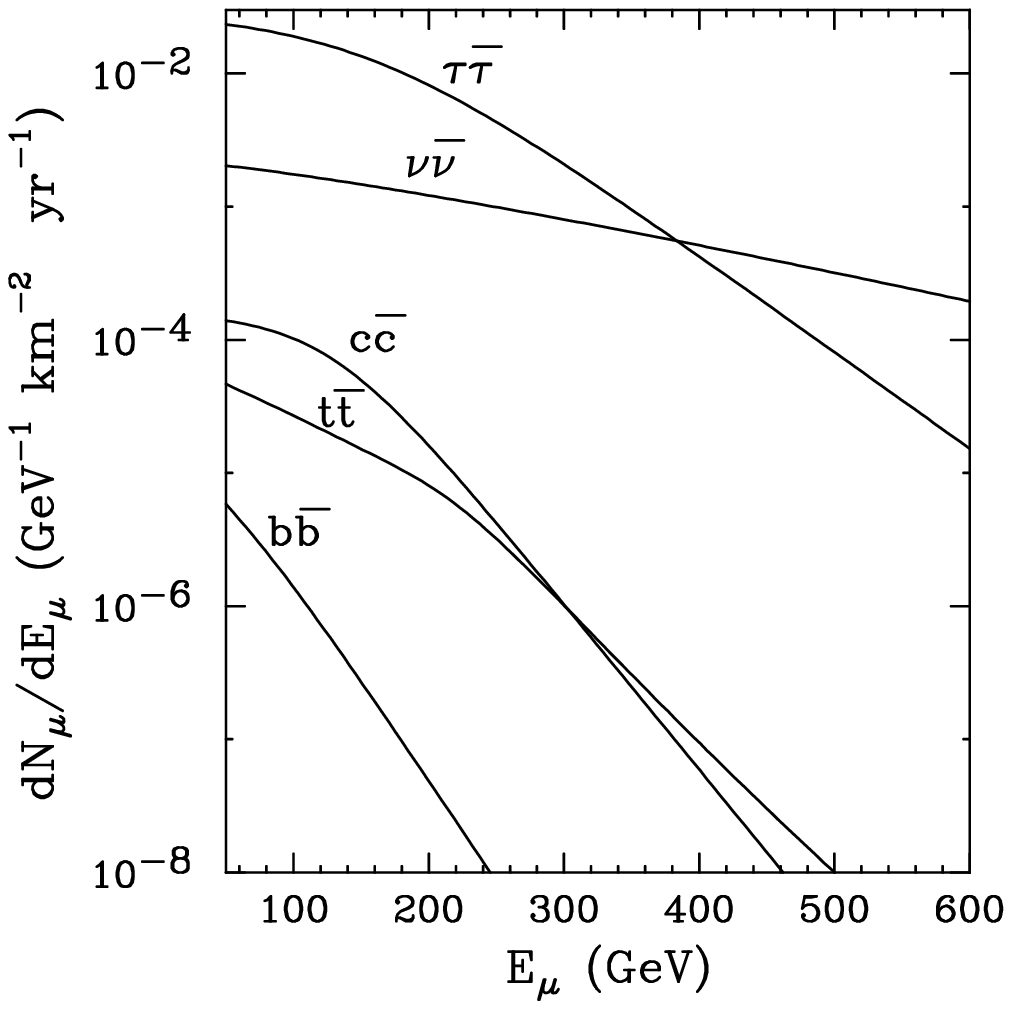}
\hfill
\includegraphics[width=3.2in]{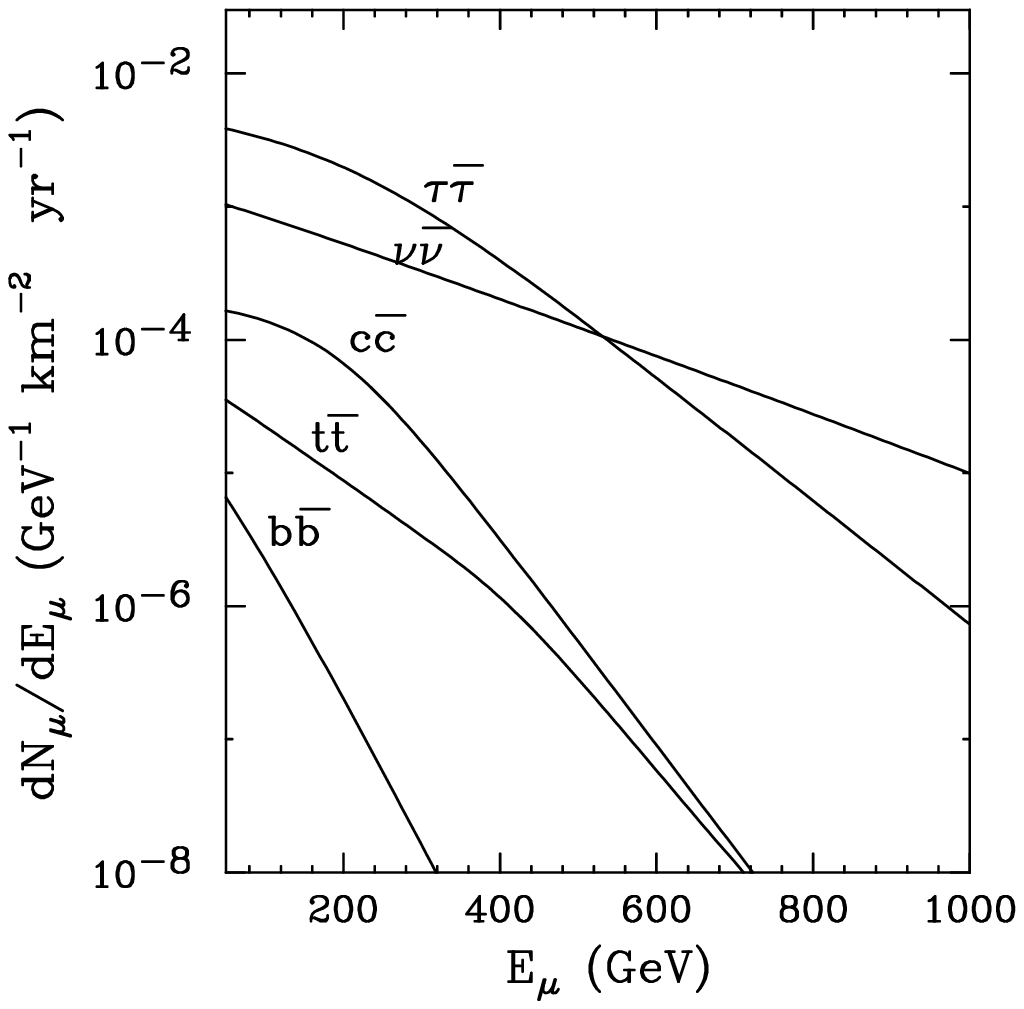}}
\caption{The spectrum of muons at the Earth generated in charged-current 
interactions of muon neutrinos generated in the annihilation of 
$600$ GeV (left side) and $1000$ GeV (right side)
dark matter particles.  The elastic scattering cross section used for 
capture in the Sun was fixed at $10^{-6}$ pb for both graphs.  
The rates are proportional to that cross section.  
We used the branching ratios given in Table~\ref{BR-table}.}
\label{fig1}
\end{figure}

In Fig.~\ref{fig1}
we show the muon flux at the surface of the Earth from $\B$
annihilations in the Sun for $m_{\B} = 600$ and $1000$ GeV, respectively.  
For the purposes of comparison, the spin-dependent cross section 
was fixed at $\sigma_{\mathrm{H,SD}} = 10^{-6}$ pb for both masses.  
The majority of events result from the indirect source 
$\B\B \ra \tau^+\tau^-$, when the $\tau$'s decay producing 
a muon neutrino either directly or through flavor oscillations.

\begin{figure}[t]
\centering\leavevmode
\includegraphics[width=3.5in]{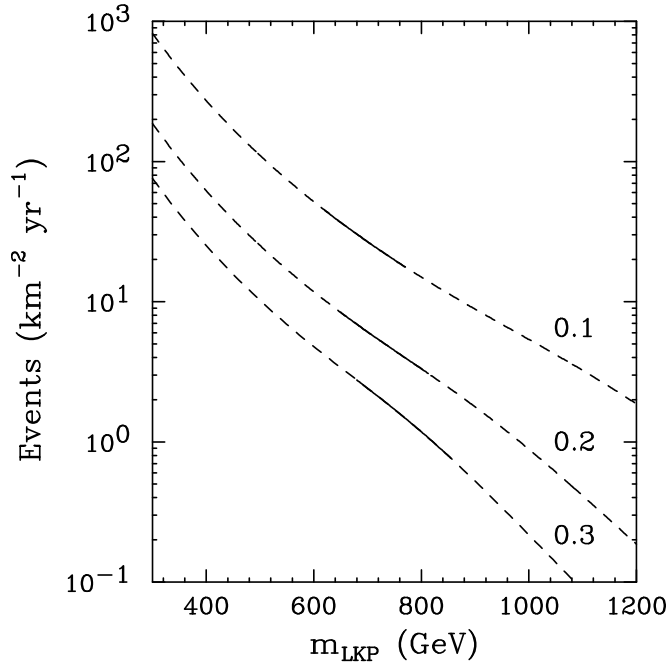}
\caption{The number of events per year in a detector with effective 
area equal to one square kilometer.  Contours are shown for 
$r_{\qR}=0.1$, $0.2$, and $0.3$.  The $r_{\qR}=0.3$
is shown merely for comparison, since this mass ratio is 
larger than would be expected from the one-loop radiative correction 
calculations of the KK mode masses.
The relic density of the $\B$'s lies within the favored range
$\Omega_{\B} h^2 = 0.16 \pm 0.04$ for the solid sections 
of each line.  The relic density is smaller (larger) for 
smaller (larger) LKP masses \cite{ST}.}
\label{fig2}
\end{figure}

Using the neutrino energy spectrum, the event rate expected at an 
existing or future neutrino telescope can be calculated.  
This is shown in Fig.~\ref{fig2} 
for a detector with an effective area of $1$ km$^2$, such as IceCube.  
Each line corresponds to a 
different value of $r_{\qR}$.  We should emphasize that the 
expected size of the one-loop radiative corrections from 
(\ref{uR-eq})--(\ref{dR-eq}) predict $0.1 \lsim r_{\qR} \lsim 0.2$ 
for $10/R \lsim \Lambda \lsim 100/R$.  For this range,
a kilometer scale neutrino telescope would be sensitive to a 
$\B$ with mass up to about 1 TeV\@.  The relic density
of the $\B$ varies from low to high values from 
left to right in the graph.  The range of mass of the $\B$ that
gives the appropriate relic density $\Omega_{\B} h^2 = 0.14 \pm 0.04$
to be dark matter was estimated from \cite{ST} and shown in the
figure by the solid sections of the lines.  
Note that coannihilation with KK quarks is not included in Fig.~\ref{fig2},
however this is not likely to significantly affect our results 
for $r_{\qR} \gsim 0.1$ as shown in the figure \cite{ST}.  
Generally the largest uncertainties in our calculations arise from 
astrophysical sources, such as the local dark matter halo density,
which we estimate to be within about a factor of two.
Combining the expected
size of the one-loop radiative corrections with a relic density
appropriate for dark matter, we find that IceCube should see 
between a few to tens of events per year.  

For detectors with smaller effective areas one simply has to scale the 
curves down by a factor $A/(1 \; \mathrm{km}^2)$ to obtain the event rate.  
In particular, for the first generation neutrino telescopes including AMANDA, 
ANTARES, and NESTOR, with effective areas of order $0.1$ km$^2$,
the event rate could be as high as a few events per year for a
$\B$ mass at the lower end of the solid line region.

A discovery of WIMP annihilations in the Sun would inevitably require
a careful analysis of the signal over background.  
The background for this class of experiments consists of atmospheric
neutrinos \cite{atmback} and neutrinos generated in cosmic ray
interactions in the Sun's corona \cite{sunback}.  In the direction of the
Sun (up to the angular resolution of a neutrino telescope), tens of events
above 100 GeV and on the order of 1 event per year above 1 TeV, per square
kilometer are expected from the atmospheric neutrino flux.  Fortunately,
for a very large volume detector with sufficient statistics, this
background is expected to be significantly reduced, and possibly 
eliminated.  Furthermore, this rate could be estimated based on the 
rate from atmospheric neutrinos, a level of about a few events per year. 
The final background is then further reduced by selecting on judiciously
chosen angular and/or energy bins.  Neutrinos generated by cosmic ray 
interactions in the Sun's corona, however, cannot be reduced in this way.
This irreducible background is predicted to be less than a few events per
year per square kilometer above 100 GeV\@.

\section{Conclusions}

The prospects for indirect detection of $\B$ dark matter is very 
promising at kilometer scale neutrino telescopes.  Using the one-loop 
radiative corrections to the KK mass spectrum, we showed that if the $\B$
lies in the mass range in which it has an acceptable present-day 
relic density to be the dark matter of the universe, then a
$1$ km$^2$ neutrino telescope is expected to detect between 
a few to tens of $\B$ annihilation events in the Sun per year.  
This mass range of the $\B$ is between about $600$ to $800$ GeV where 
coannihilations with right-handed KK leptons plays a significant
role.  The relatively large signal relies on the Sun reaching
equilibrium between $\B$ capture and annihilation, which we explicitly
verified for this range of $\B$ masses.  

Although we have focused on the $\B$ mass range that results in 
the appropriate dark matter relic density as currently consistent
with cosmology, it is straightforward to extrapolate to other masses.
In fact, there are two particle physics effects that are expected to 
lead to a \emph{lowering} of the $\B$ mass for a fixed relic density.
The first effect is the inclusion of coannihilation with left-handed KK
leptons.  The second effect, in a six dimensional model, is that 
there are typically multiple LKPs corresponding to multiple conserved 
parities.  In both of these cases the lowering of the mass of the LKP(s)
for a fixed relic density implies a larger event rate at neutrino 
telescopes.  We are therefore optimistic that future detectors will 
find or exclude this fascinating possibility.

\section*{Acknowledgments}

We thank G\'eraldine Servant and Tim Tait for several valuable
discussions and for reading the manuscript.  
We also thank John Beacom for informing us of Ref.~\cite{crotty}
and for enlightening discussions concerning its effects on our results.
G.D.K.\ thanks the Aspen Center for Physics and the T-8 group
of Los Alamos National Laboratory for hospitality where part of this 
work was completed.  This work was supported in part by a DOE grant 
No.\ DE-FG02-95ER40896 and in part by the Wisconsin Alumni Research 
Foundation.


\begin{thebibliography}{99}
\singlespaced

\bibitem{cmb}
P.~de Bernardis {\it et al.}  [Boomerang Collaboration],
Nature {\bf 404}, 955 (2000)
[arXiv:astro-ph/0004404];
S.~Hanany {\it et al.},
Astrophys.\ J.\  {\bf 545}, L5 (2000)
[arXiv:astro-ph/0005123];
A.~Balbi {\it et al.},
Astrophys.\ J.\  {\bf 545}, L1 (2000)
[Erratum-ibid.\  {\bf 558}, L145 (2001)]
[arXiv:astro-ph/0005124].
C.~B.~Netterfield {\it et al.}  [Boomerang Collaboration],
Astrophys.\ J.\  {\bf 571}, 604 (2002)
[arXiv:astro-ph/0104460].
C.~Pryke, {\it et al.},
Astrophys.\ J.\  {\bf 568}, 46 (2002)
[arXiv:astro-ph/0104490];

\bibitem{sn}
S.~Perlmutter {\it et al.}  [Supernova Cosmology Project Collaboration],
Astrophys.\ J.\  {\bf 517}, 565 (1999)
[arXiv:astro-ph/9812133].

\bibitem{rotation}
K.G.~Begeman, A.H.~Broeils, R.H.~Sanders,
Mon. Not. R. Astr. Soc. {\bf 249}, 523 (1991).

\bibitem{bbn}
S.~Burles, K.~M.~Nollett, J.~N.~Truran and M.~S.~Turner,
Phys.\ Rev.\ Lett.\  {\bf 82}, 4176 (1999)
[arXiv:astro-ph/9901157].

\bibitem{luminous}
M.~Fukugita, C.~J.~Hogan and P.~J.~Peebles, 
Astrophys.~J. {\bf 503}, 518 (1998) 
[arXiv:astro-ph/9712020].

\bibitem{structure}
M.~Davis, G.~Efstathiou, C.~S.~Frenk and S.~D.~White,
Astrophys.\ J.\  {\bf 292}, 371 (1985).

\bibitem{KolbSlansky}
E.~W.~Kolb and R.~Slansky,
Phys.\ Lett.\ B {\bf 135}, 378 (1984).

\bibitem{DDG2}
K.~R.~Dienes, E.~Dudas and T.~Gherghetta,
Nucl.\ Phys.\ B {\bf 537}, 47 (1999)
[arXiv:hep-ph/9806292].

\bibitem{ACD}
T.~Appelquist, H.~C.~Cheng and B.~A.~Dobrescu,
Phys.\ Rev.\ D {\bf 64}, 035002 (2001)
[arXiv:hep-ph/0012100].

\bibitem{antoniadis}
I.~Antoniadis,
Phys.\ Lett.\ B {\bf 246}, 377 (1990);
I.~Antoniadis, K.~Benakli and M.~Quiros,
Phys.\ Lett.\ B {\bf 331}, 313 (1994).

\bibitem{CMS}
H.~C.~Cheng, K.~T.~Matchev and M.~Schmaltz,
arXiv:hep-ph/0204342;
arXiv:hep-ph/0205314.

\bibitem{ST}
G.~Servant and T.~M.~Tait,
arXiv:hep-ph/0206071.

\bibitem{CFM}
H.~C.~Cheng, J.~L.~Feng and K.~T.~Matchev,
arXiv:hep-ph/0207125.

\bibitem{DP}
B.~A.~Dobrescu and E.~Poppitz,
Phys.\ Rev.\ Lett.\  {\bf 87}, 031801 (2001)
[arXiv:hep-ph/0102010].

\bibitem{ADPY}
T.~Appelquist, B.~A.~Dobrescu, E.~Ponton and H.~U.~Yee,
Phys.\ Rev.\ Lett.\  {\bf 87}, 181802 (2001)
[arXiv:hep-ph/0107056].

\bibitem{GGH}
H.~Georgi, A.~K.~Grant and G.~Hailu,
Phys.\ Lett.\ B {\bf 506}, 207 (2001)
[arXiv:hep-ph/0012379].

\bibitem{susydm}
J.~L.~Feng, K.~T.~Matchev and F.~Wilczek,
Phys.\ Rev.\ D {\bf 63}, 045024 (2001)
[arXiv:astro-ph/0008115];
V.~D.~Barger, F.~Halzen, D.~Hooper and C.~Kao,
Phys.\ Rev.\ D {\bf 65}, 075022 (2002)
[arXiv:hep-ph/0105182];
V.~Bertin, E.~Nezri and J.~Orloff,
arXiv:hep-ph/0204135.

\bibitem{capture}
A.~Gould, Astrophys.\ J.\ {\bf 388}, 338 (1991); 
G.~Jungman, M.~Kamionkowski and K.~Griest,
Phys.\ Rept.\  {\bf 267}, 195 (1996)
[arXiv:hep-ph/9506380].

\bibitem{spinfraction}
G.~K.~Mallot,
in {\it Proc. of the 19th Intl. Symp. on Photon and Lepton Interactions at High Energy LP99 } ed. J.A. Jaros and M.E. Peskin,
Int.\ J.\ Mod.\ Phys.\ A {\bf 15S1}, 521 (2000)
[eConf {\bf C990809}, 521 (2000)]
[arXiv:hep-ex/9912040];
W.~M.~Alberico, S.~M.~Bilenky and C.~Maieron,
Phys.\ Rept.\  {\bf 358}, 227 (2002)
[arXiv:hep-ph/0102269].

\bibitem{equ}
K.~Griest and D.~Seckel,
Nucl.\ Phys.\ B {\bf 283}, 681 (1987)
[Erratum-ibid.\ B {\bf 296}, 1034 (1988)];
A.~Gould,
Astrophys.\ J.\  {\bf 321}, 571 (1987).

\bibitem{edsjo}
J.~Edsjo,
Nucl.\ Phys.\ Proc.\ Suppl.\  {\bf 43}, 265 (1995)
[arXiv:hep-ph/9504205].

\bibitem{JK}
G.~Jungman and M.~Kamionkowski,
Phys.\ Rev.\ D {\bf 51}, 328 (1995)
[arXiv:hep-ph/9407351].

\bibitem{crotty}
P.~Crotty,
arXiv:hep-ph/0205116.

\bibitem{review}
F.~Halzen and D.~Hooper,
Rept.\ Prog.\ Phys.\  {\bf 65}, 1025 (2002)
[arXiv:astro-ph/0204527].
E.~Andres {\it et al.},
Nature {\bf 410}, 441 (2001).

\bibitem{crosssection}
R.~Gandhi, C.~Quigg, M.~H.~Reno and I.~Sarcevic,
Astropart.\ Phys.\  {\bf 5}, 81 (1996)
[arXiv:hep-ph/9512364].


\bibitem{range}
S.~I.~Dutta, M.~H.~Reno, I.~Sarcevic and D.~Seckel,
Phys.\ Rev.\ D {\bf 63}, 094020 (2001)
[arXiv:hep-ph/0012350].

\bibitem{atmback}
T.~K.~Gaisser, F.~Halzen and T.~Stanev,
Phys.\ Rept.\  {\bf 258}, 173 (1995)
[Erratum-ibid.\  {\bf 271}, 355 (1996)]
[arXiv:hep-ph/9410384].

\bibitem{sunback}
L.~Bergstrom, J.~Edsjo and P.~Gondolo,
Phys.\ Rev.\ D {\bf 55}, 1765 (1997)
[arXiv:hep-ph/9607237];
L.~Bergstrom, J.~Edsjo and P.~Gondolo,
Phys.\ Rev.\ D {\bf 58}, 103519 (1998)
[arXiv:hep-ph/9806293].

\end{thebibliography}
\end{document}